\documentclass[twocolumn]{emulateapj}
\begin{document}
\shorttitle{Variability in magnetized $\nu$-cooled accretion disks}

\shortauthors{Carballido \& Lee}

\title{Characterizing the time variability in magnetized
  neutrino--cooled accretion disks: signatures of the gamma-ray burst
  central engine}

\author{Augusto Carballido and William H. Lee} \affil{Instituto de
  Astronom\'{\i}a, Universidad Nacional Aut\'{o}noma de M\'{e}xico
  \\ Apdo. Postal 70-543, M\'{e}xico D.F. 04510, MEXICO}
  
\begin{abstract}
The central engine of Gamma Ray Bursts is hidden from direct probing
with photons mainly due to the high densities involved. Inferences on
their properties are thus made from their cosmological setting,
energetics, low-energy counterparts and variability. If GRBs are
powered by hypercritical accretion onto compact objects, on small
spatial scales the flow will exhibit fluctuations, which could in
principle be reflected in the power output of the central engine and
ultimately in the high energy prompt emission. Here we address this
issue by characterizing the variability in neutrino cooled accretion
flows through local shearing box simulations with magnetic fields, and
then convolving them on a global scale with large scale dynamical
simulations of accretion disks. The resulting signature is
characteristic, and sensitive to the details of the cooling mechanism,
providing in principle a discriminant for GRB central engine
properties.
\end{abstract}


\section{Introduction}\label{sec:intro}

Gamma Ray Bursts (GRBs) are thought to be driven by mass accretion
onto compact objects \citep{wb06,nakar07,lrr07,grrf09}. The enabling
cooling mechanism allowing accretion is neutrino emission, raising the
usual Eddington rate for photons by sixteen orders of magnitude
\citep{pwf99,npk01,km02,bel03,dmpn02,setiawan04,lrrp05,cb07}. This
applies both to long and short GRBs, perhaps due to the collapse of
massive rotating stars \citep{woosley93,mw99} and compact binary
mergers \citep{eichler89,bp91,narayan92}, respectively, or magnetized
neutron stars \citep{usov92}. The prompt gamma-ray emission originates
at $\simeq 10^{14}-10^{16}$~cm from this source, possibly from
internal shocks in the relativistic outflow generated by the engine
\citep{zhang04}. The engine itself is hidden from view due to the high
opacities, and can be potentially probed directly only through
gravitational waves and neutrinos.

The observed time series in GRB prompt emission show diversity between
events \citep[see, e.g.,][]{norris96}: some have a single peak, others
multiple emission episodes, with correlations between the fluence of
the active period and the length of the quiescent interval preceding
it \citep{rr01a}. On top of this, rapid (ms) fluctuations are
routinely observed. Fourier analysis of the high-energy light curves
of the prompt emission in GRBs in the source frame reveals power-law
spectra \citep{bel98,bel00,ryde03}, with index $\simeq -5/3$ and a
break at $\simeq 1-2$~Hz.  Variability is likely due to a combination
of several effects, allowing in principle an additional way to probe
the central engine indirectly. Some are probably intrinsic to the
progenitor: the distribution of angular momentum with radius inside
the star in the case of a collapsar may lead to distinct episodes of
energy release \citep{kumar08,perna10,lclrr10}; the fall back at late
times of material stripped from a tidally disrupted neutron star is
capable of powering secondary accretion episodes
\citep{rosswog07,lrrlc09}; hydrodynamical or magnetic instabilities in
the accretion disk may result in intermittent accretion
\citep{perna06,proga06,taylor10}. Others can come from the
relativistic outflow: the interaction of a jet with high Lorentz
factor with the stellar envelope before breakout can lead to
irregularities and shocking \citep{morsony10}; the outcome of internal
shocks between shells in the flow depends on the variation in mass and
energy upon ejection \citep{panaitescu99,rr01b,bosnjak09,mendoza09}.
An additional factor, upon which we focus here, is related to the
variability present in the accretion disk as a result of turbulent
motions. The dissipation is related to the local hydrodynamical
variables, and as these vary in time, so will the energy output.

The magnetorotational instability (MRI) \citep{bh98} is a possible
mechanism that will allow for the transport of angular momentum in
accretion disks with differential rotation. Its behavior under the
physical conditions in neutrino cooled disks, which are similar to
those occurring in supernovae, but different from the usual ones
present in X-ray binaries and AGN, has come under scrutiny more
recently \citep{thompson05,masada07,rossi08,obergaulinger09}.  One
such difference lies in the sensitivity of the cooling rate to
temperature and is at the heart of this work. Whereas, for example,
the photon bremsstrahlung emissivity scales as $\dot{q} \propto
T^{1/2}$ in the optically thin limit, for neutrinos $\dot{q} \propto
T^{\beta}$, where $\beta \simeq 6-9$ depending on the cooling
process. Further, while photon-cooled disks are typically optically
thick, $\tau_{\gamma} \gg 1$, for a wide range of relevant parameters
their neutrino-cooled counterparts are optically thin, $\tau_{\nu}
\leq 1$, tightly coupling the local conditions to the emitted
luminosity.

In this {\em Letter}, we characterize the local variability in
neutrino cooled disks through shearing box MHD simulations
(\S~\ref{sec:shear}), and then use the results of global disk
simulations to scale the results for the central engine as a whole
(\S~\ref{sec:global}). Prospects for placing constraints on GRBs are
discussed in \S~\ref{sec:prospects}.

\section{The small scale: numerics and physics in the shearing box calculations}\label{sec:shear}

The local flow in the disk is modeled by the shearing box
approximation \citep{hawley95}: a rectangular Cartesian coordinate
system represents a local neighborhood inside the disk, at an
arbitrary orbital radius $R_{0}$ with dimensions which are much
smaller than $R_{0}$. The $x$, $y$ and $z$ axes represent the radial,
azimuthal and vertical directions in the disk, respectively. The
radial component of the central object's gravity is included, and
differential rotation is replaced by a Keplerian shear flow along the
$y$ direction. Periodic boundary conditions are set in $y$ and $z$,
while in $x$ these are ``shearing periodic'': upon crossing a radial
boundary, a fluid element is displaced along $y$ by an amount given by
the shear value. The Eulerian ZEUS-3D code [\citet{stone92a,stone92b}]
is used to solve the equations of ideal MHD. These are:
 
 \begin{equation}
 \frac{\partial \rho}{\partial t}+\nabla\cdot\left(\rho\mathbf{v}\right)=0,
 \end{equation}
  
 \begin{equation}
 \frac{\partial \textbf{v}}{\partial t}+\textbf{v}\cdot\nabla \textbf{v} =
 -\frac{1}{\rho}\nabla \left(P+\frac{B^{2}}{8\pi}\right)
 +\frac{\left(\textbf{B}\cdot\nabla\right)\textbf{B}}{4\pi\rho}-2\mathbf{\Omega_{0}}\times\mathbf{v}
 +\frac{3}{2}\Omega_{0}^{2}x\hat{\textbf{x}},
 \end{equation}
 
 \begin{equation}
 \frac{\partial\textbf{B}}{\partial t}=\nabla \times\left(\textbf{v}\times\textbf{B}\right),
 \end{equation}
 
 \begin{equation}\label{eq:energy}
 \frac{\partial  e}{\partial t}=-P\nabla\cdot\mathbf{v} -C\rho T^{\beta},
 \end{equation}
 
\noindent
and an ideal gas equation of state with $P=(\gamma-1) e$ where
$\mathbf{v}$, $\rho$ and $P$ are the gas velocity, density and
pressure, respectively, $\mathbf{B}$ is the magnetic field,
$\mathbf{\Omega_{0}}=(0,0,\Omega_{0})$ is the disk angular frequency,
$e$ is the internal energy density, and $\hat{\mathbf{x}}$ is a
unitary radial vector.

The second term on the right-hand side of Eq.(\ref{eq:energy}) is
added to model neutrino cooling processes, with the nature of the
particular emission mechanism fixing the value of $\beta$. In GRB
central engines, the dominant processes are $e^{\pm}$ pair
annihilation \citep{itoh96}, with $\dot{q}_{\rm ann} \propto T^{9}$,
and $e^{\pm}$ capture onto free neutrons and protons \citep{lmp01},
with $\dot{q}_{\rm cap}\propto T^{6}$. A fraction $\zeta$ of the
initial internal energy of the fluid is removed over 100 orbital
periods at $R_{0}$, the duration of the simulation. We used
$\zeta=0.1$, thus the cooling time is much longer than the dynamical
time.

Initially the density, internal energy and pressure in the box are
uniform, and the magnetic field is vertical with non-zero net magnetic
flux across the box. The ratio of gas to magnetic pressure is $P_{\rm
  gas}/P_{\rm mag}=400$. The velocity field is randomly perturbed with
amplitude $v_{\rm pert}=10^{-3}c_{\rm s}$, where $c_{\rm s}$ is the
sound speed. We have performed simulations at varying resolution to
test for convergence (the highest having 128x256x128 zones in $x,y,z$,
respectively). From the power spectrum of the velocity field
$\hat{v}_{\rm i}(k_{\rm i})$, we have verified that our results are
converged and unaffected by resolution changes at the level
reported. Turbulent motions induced by the MRI develop in the
simulation on a time scale $t_{\rm MRI}\propto 1/\Omega_{0}$, leading
to turbulent transport of angular momentum. The fundamental output of
the simulations is the luminosity as a function of time of the
shearing box, which we denote as $s(t)$.

\begin{figure}[h!]
  \begin{center}
  \epsscale{1.0}    
\includegraphics[width=\columnwidth,angle=0,scale=1.]{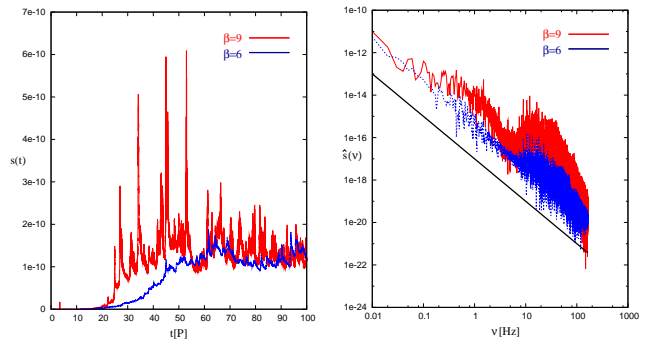}
    \caption{Left: Energy output (arbitrary units) as a function of
      time, $s(t)$, for shearing box simulations with cooling
      mimicking pair capture (blue, $\beta=6$) and pair annihilation
      (red, $\beta=9$).  The time is given in units of the orbital
      period at $R_{0}$, the fiducial shearing box radial
      position. Right: Fourier transforms $\hat{s}(\nu)$ of
      $s(t)$. The reference power law has index -2.}
    \label{fig:luminosity}
  \end{center}
\end{figure} 

Luminosity variations for $s(t)$ are shown in
Figure~\ref{fig:luminosity} for $\beta=6, 9$. Rapid variability is
present, reflecting the turbulent motions in the box. These changes
can be characterized by their Fourier transform, $\hat{s}(\nu)$, where
$\nu$ is the signal frequency (Figure~\ref{fig:luminosity}). Roughly,
the spectrum is a power law, $\hat{s}(\nu) \propto \nu^{-2}$, over a
wide range of frequencies in the case of $\beta=6$, with more
complicated behavior superimposed in the case of $\beta=9$. It is
clear that the variations in energy release have a characteristic
behavior, with the details varying according to the specific cooling
mechanism.

\section{The large scale: extension to global disk variability}\label{sec:global}

To find the global variability we require the superposition of a very
large number of smaller regions represented by the shearing box
calculations, and approximations are necessary to proceed further. We
assume equatorial and azimuthal symmetry, and also take the disk to be
in a stationary state. Obtaining global quantities thus requires
integrating over the polar angle $\phi \in [0,2\pi]$ and radius $r \in
[r_{\rm in},\infty]$. We are still faced with the superposition of a
potentially large number of zones, each uncorrelated with the other,
and which could destroy any coherent signature of the processes
occuring on small scales. A natural quantity in the disk, however,
enforces causality for a given frequency: the local sound speed,
$c_{\rm s}$.

Consider a disk annulus at radius $R$, with sound speed $c_{\rm
  s}$. The correlation length over which signals with frequency $\nu$
propagate is $l_{\nu}=c_{\rm s}/\nu$. The azimuthal velocity is
$v_{\phi}=R\Omega$, and from vertical hydrostatic balance, the
pressure scale height $H$ satisfies $H/R\simeq c_{\rm
  s}/v_{\phi}$. Thus $c_{\rm s}= 2 \pi R f (H/R)$, with $f$ being the
orbital frequency at radius $R$. In neutrino cooled accretion disks,
$A=H/R\approx $~cst \citep{lrrp05,lrr07}, so that $l_{\nu}=2 \pi R A
f/\nu$. The number of distinct azimuthal zones within the ring for
frequency $\nu$ is $N_{\phi, \, \nu}=2 \pi R/l_{\nu}=\nu/(A
f)$. Dynamical simulations of such disks show that $A \approx
1/10$. If $N_{\phi, \, \nu} \geq 1$, the signal at frequency $\nu$ is
thus diminished with respect to the idealized case of entirely
in-phase contributions by a factor $N_{\phi \, \nu}^{1/2}$, due to the
random nature of the phase de-correlation. If, on the contrary,
$N_{\phi \, \nu}\leq 1$, the signal at frequency $\nu$ is added
constructively. This acts as a broad low-pass filter by slightly
suppressing the signal at high frequencies, when $\nu \gg A
f$. Carrying out this filtering procedure also requires scaling the
fiducial model to different radii, which we now address.  \\

The emissivity is not a function of azimuthal position, but it is one
of radius. Given the extreme sensitivity of the cooling rate to
temperature, only the innermost regions of the disk contribute to the
luminosity, and thus to its variability properties. The argument given
above for the decoupling of azimuthal zones in the disk for a
particular frequency is also applicable to the radial
coordinate. Advection may modify this slightly, but for our present
purpose it is valid enough to say that as the flow is split up into
$N_{\phi\, \nu}$ zones in azimuth, it is divided into rings of
characteristic size $l_{\nu}$. The number of radial zones in the disk
at frequency $\nu$ is $N_{R \, \nu}=r_{\rm out}/l_{\nu}$, where
$r_{\rm out}$ is the outer radius of the disk. Using the previous
results, $N_{R \, \nu}=r_{\rm out} \nu/(2 \pi R A f)$. A second
low-pass filter needs to be considered in analogy with the azimuthal
one, and the signal is now suppressed by a factor $N_{R \, \nu}^{1/2}$
if $N_{R \, \nu} \geq 1$.

\begin{figure}[h!]
  \begin{center}
  \epsscale{1.0}
  \includegraphics[width=\columnwidth,angle=0,scale=1.]{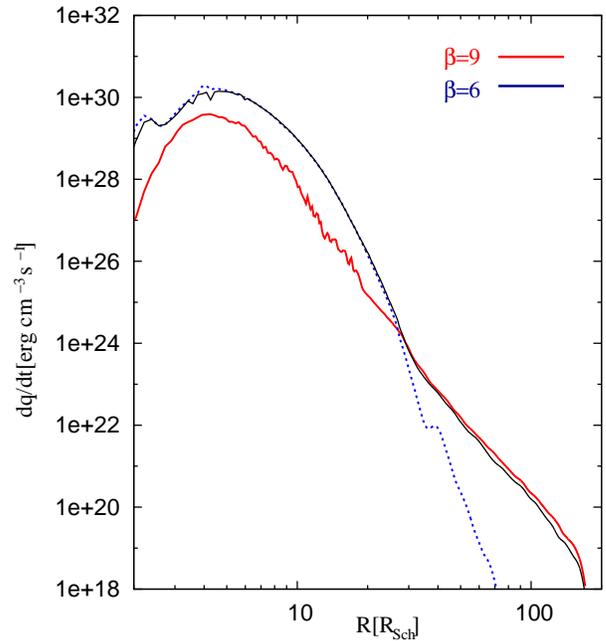}
    \caption{Equatorial ($z=0$) neutrino emissivities,
      $\dot{q}(R/R_{\rm Sch})$, $R_{\rm Sch}=2GM_{\rm BH}/c^{2}$, for
      two dimensional, azimuthally symmetric global dynamical
      simulations of accretion disks around a black hole with
      $M=4M_{\odot}$. The blue (red) line corresponds to simulation
      $T6$ ($T9$), where only pair captures (annihilation) were used
      to compute the cooling rate. The solid black line is the result
      of a simulation including both processes. }
    \label{fig:qdotR}
  \end{center}
\end{figure}

In order to carry out this procedure, we have used the results of
global, 2D simulations of neutrino-cooled accretion disks
\citep{lrrp05,lrr07} orbiting a stellar mass black hole, evolved in
the pseudo-relativistic potential of \citet{pw80}. For consistency
purposes, we have re-computed their evolution for this work for two
test cases, with $M_{\rm BH}=4M_{\odot}$ and $q=M_{\rm disk}/M_{\rm
  BH}=7.5 \times 10^{-3}$. In the first model, $T6$, we have
eliminated all sources of cooling except from pair captures, where
$\dot{q}\propto T^{6}$, while in the second, $T9$, we have only kept
that due to pair annihilation, with $\dot{q} \propto T^{9}$. Each
model can then be more faithfully compared with the two fiducial runs
for the shearing box with $\beta=6, 9$ respectively, using a
near-stationary state where the accretion rate is
$\dot{M}=0.05$~M$_{\odot}$~s$^{-1}$.

We first compute the equatorial radial profile of emissivity in
simulations $T6$ and $T9$, $\dot{q}(R)$, shown in
Figure~\ref{fig:qdotR}. At small radii the dominant contribution comes
from pair capture, while annihilation dominates for $R \geq R^{*}=30
R_{\rm Sch}$. The combination is a strongly decreasing function of
$R$, so the radial convolution need not be extended to infinity, and
pair capture dominates the total emission, as well as its general
behavior.  The inner radius of the disk is $r_{\rm in}=r_{\rm ms}=3
R_{\rm Sch}$, where the effects of General Relativity truncate the
flow. We have used $r_{\rm out}=85 R_{\rm Sch}$, and checked that the
final result is insensitive to this choice, as long as $r_{\rm out}
\gg R^{*}$.

To perform the radial integral, the time series $s(t)$ and
$\hat{s}(\nu)$ need to be scaled to the disk at different radii
$R$. The characteristic break frequency in the fiducial spectrum for
the volume-averaged total (Maxwell and Reynolds) stress, measured as
an equivalent $\alpha$ viscosity parameter, is roughly at the orbital
frequency, $f$. This is related to the nature of the MRI, where,
$t_{\rm MRI} \simeq 1/\Omega$. The orbital angular frequency is known
as a function of radius, $\Omega=(GM_{\rm BH}/R)^{1/2}(R/(R-R_{\rm
  Sch}))$, so we can convert from frequency to radius, $\hat{s}(\nu)$
to $\hat{s}(R)$ and integrate over different annuli (the expression
for $\Omega$ is appropriate for the pseudo-relativistic potential
used).

\begin{figure}[h!]
  \begin{center}
  \epsscale{1.0} 
\includegraphics[width=\columnwidth,angle=0,scale=1.]{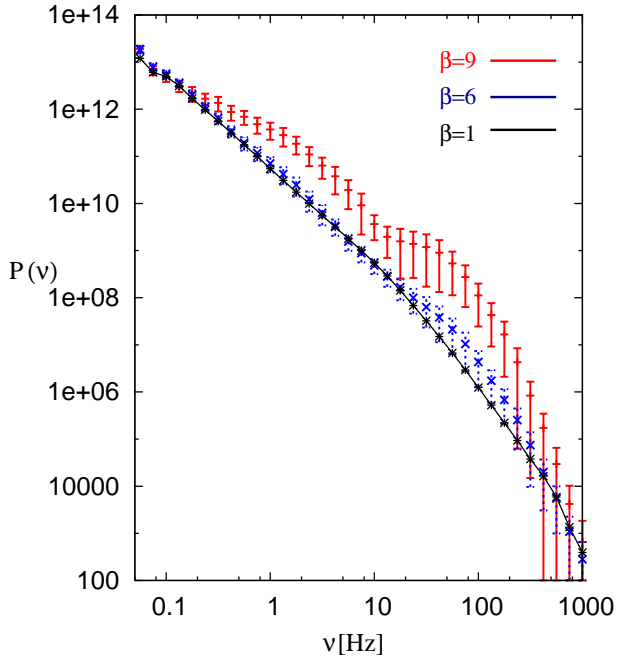}    
    \caption{Convolved power spectra over the entire disk,
      Eq.~\ref{eq:pds}, for models $T6$ (red) and $T9$ (blue). The
      power is given in arbitrary units as a function of frequency (in
      Hz) for a black hole with $4M_{\odot}$. The black line shows the
      result of a test calculation with local cooling in the shearing
      box simulation set to $\dot{q} \propto T$, i.e., $\beta=1$. The
      error bars for $\beta=6, 9$ include the scatter from the Fourier
      transform in the plotted frequency bins and sampling errors in
      the radial discretization of the convolution.}
    \label{fig:pds}
  \end{center}
\end{figure}

The power associated to the variability in the neutrino luminosity can
now be computed as:
\begin{equation}\label{eq:pds}
P(\nu)=\int_{r_{\rm rin}}^{r_{\rm out}} \dot{q}(R) \, \Phi(N_{\phi \,
  \nu}) \, \rho(N_{R \, \nu})\, \hat{s}(R) \, 2 \pi R dR,
\end{equation}
where $\dot{q}(R)$ corresponds to model $T6$ or $T9$, and
$\Phi(N_{\phi \, \nu})$ and $\rho(N_{R \, \nu})$ account for the
filtering at high frequencies. Figure~\ref{fig:pds} shows the
resulting spectra, where several features clearly stand out. At low
frequencies, $\nu \leq \nu_{0} \simeq $0.1~Hz, the spectrum is the
same in both cases, $P(\nu) \propto \nu^{-2}$.  For $\nu \geq \nu_{0}$
the results diverge: model $T6$ approximately maintains the power law
decay, while model $T9$ shows an excess, and in general more complex
behavior, with broad features at $\approx 2$ and 40~Hz overlaid on a
decay with lower index, $P(\nu)\propto \nu^{-1.7}$. At $\nu=\nu_{1}
\simeq$ 100~Hz, a break to a more rapidly decaying power law
terminates both spectra. The errors are greater for model $T9$, due to
greater scatter in the Fourier transform. The thin black line in
Figure~\ref{fig:pds} is from a shearing box test calculation with
$\dot{q} \propto T$, i.e., with $\beta=1$ (the smaller error bars are
not plotted for clarity). As the sensitivity of the cooling term to
the temperature rises, the trend is for an excess in power at higher
frequencies, and a more moderate background power law decay.

\section{Summary, discussion and prospects for observability in GRBs}\label{sec:prospects}

Through shearing box MHD simulations we have characterized the time
variability of the energy release at small scales in a neutrino-cooled
accretion disk around a black hole, where energy losses primordially
come from $e^{\pm}$ pair capture onto free nucleons and protons and
pair annihilation. With the use of cooling profiles from large scale,
two-dimensional simulations of full disks, we have convolved this
local variability to obtain a global signature of time variations in
the power output, through the power density spectrum of the neutrino
luminosity. Since accretion is enabled by the cooling through
neutrinos, we take this as an indicator of central engine variability
which will be reflected in the relativistic outflow eventually giving
rise to a GRB.

The power spectrum exhibits characteristic features related to the
general nature of the neutrino cooling in the optically thin regime,
and particularly to its temperature dependence. A background power law
decay, with index $\simeq 1.7-2$ extends approximately from
0.1-100~Hz. For emissivities $\dot{q}\propto T^{\beta}$, additional
power appears at high frequencies as $\beta$ rises, with the
particular values of the slopes and cuttoffs scaling with the cooling
mechanism (Fig.\ref{fig:pds}). This result thus provides in principle
a discriminant for GRB central engines powered by neutrino-cooled
accretion flows, and illustrates how any local mechanism can be used
in the same way to test its viability.

The reason for the difference when modifying the cooling mechanism is
fundamentally related to the local energy balance. For $\dot{q}\propto
T^{9}$ (model $T9)$ and a given internal energy supply, $e$, the local
cooling time $t_{\rm cool} = e/\dot{q}$ is shorter under a temperature
perturbation $\Delta T$ than when $\dot{q}\propto T^{6}$ (model
$T6$). The power is thus higher at such frequencies, leading to the
observed displacement in Fig.~\ref{fig:pds}. The argument holds when
comparing model $T6$ with the test run at $\beta=1$, and gives a way
to characterize the accretion flow and discriminate between competing
mechanisms, if the variations in the neutrino luminosity are directly
reflected in the accretion power output which drives a relativistic
flow.

Two possible scenarios in which the difference between the cooling
regimes studied here may be of relevance are worthy of note. First,
the mass of the black hole introduces a scaling into the problem when
combined with neutrino cooling.  If $M_{\rm BH}$ is too large, the
density and temperature in the accretion flow can be too low for
neutrino cooling to operate.  As the BH mass is reduced, pair
annihilation, with $\dot{q}\propto T^{9}$ first becomes effective as
an energy sink, followed by pair captures, with $\dot{q}\propto
T^{6}$. A signature of the BH mass is thus in principle available in
the variability of the flow. The second case is related to the time
evolution of the flow, assuming a certain amount of mass $M_{\rm
  disk}$ is initially available for accretion, and no further feeding
of the central engine takes place. Global disk simulations show that
the density and temperature drop as the disk drains into the BH on the
viscous time scale. Given enough time, the whole disk will lie on the
branch cooled by pair annihilation where $\alpha$ particles have
formed, below $\log[\rho(\mbox{g~cm$^{-3}$})]\simeq 6.5$ and
$\log[T(\mbox{$^{\circ}$K})]\simeq 10$ for the adopted black hole mass
(see also Figure~2 in \citet{lrrlc09}). Thus the variability may
initially behave as in case $T6$, and end as in case $T9$ (the
luminosity will have decreased substantially by then, along with the
accretion rate). The associated GRB, if one occurs, need not
necessarily be powered by neutrinos in order for these effects to be
apparent. The fact that neutrinos are responsible for the cooling
process allowing accretion makes them relevant in this context.

A number of limitations apply to this study, and can be matters for
further study. First, the scaling we have used to infer the
variability at different radii is strictly valid if the flow is
adiabatic. Explicit cooling thus violates this assumption. However,
given the choice of $\zeta$ and $t_{\rm cool}$ the associated time
scales are such that $t_{\rm cool} \gg t_{\rm dyn}$, making this a
reasonable approximation. Second, we have used the neutrino luminosity
as a proxy for the manifestation of central engine activity, which
could have a neutrino component, but need not be restricted to it. For
example, magnetic fields may power relativistic outflows leading to
the observed high-energy emission. In this sense, an alternative
characterization based on the mass accretion rate $\dot{M}$ through
the disk may be of use as well, which can then be associated to the
energy output of the relativistic outflow as $L_{\rm rel} \propto
\dot{M}c^{2}$. Third, this is only the first filter any variability
originating in a disk needs to go through before emerging as
high-energy photons. A variable mass accretion rate and energy
conversion efficiency, propagation through the stellar envelope (for a
massive star progenitor), external shocks, and general relativistic
effects \citep{birkl07}, among others, can each potentially leave
their own fingerprint on the power spectrum. The background upon which
they will do so, however, must ultimately be related to the flow in
the disk itself if that is what is driving the energy release, and is
what we focus on in this work. Finally, other mechanisms can
potentially be responsible for angular momentum transport and
variability in the disk itself, such as thermal, viscous or
gravitational instabilities, none of which we have considered here.

\acknowledgments This work was supported in part through CONACyT grant
83254. AC acknowledges support a DGAPA-UNAM postdoctoral
fellowship. We thank the referee for useful criticism of the
manuscript.

\end{document}